\documentclass[aps,epsfigure,twocolumn,superscriptaddress,showkeys,nofootinbib]{revtex4-1}
\usepackage[colorlinks=true,linkcolor=blue,urlcolor=blue,citecolor=blue,pdfusetitle]{hyperref}
\usepackage[utf8]{inputenc}
\usepackage{amsmath}
\usepackage{braket}
\usepackage{graphicx}
\usepackage{amsfonts}
\usepackage{amssymb}
\usepackage{enumerate}
\usepackage{xcolor}

\begin{document}

\title{Qiskit Variational Quantum Classifier on the Pulsar Classification Problem}

\author{Anna B. M. Souza}
\email{annabiah.beatriz@gmail.com}
\affiliation{Modelagem Computacional, Universidade SENAI CIMATEC.}
\affiliation{LAQCC - Latin American Quantum Computing Center, Universidade SENAI CIMATEC.}

\author{Clebson Cruz}
\email{clebson.cruz@ufob.edu.br}
\affiliation{Centro de Ciências Exatas e das Tecnologias, Universidade Federal do Oeste da Bahia - Campus Reitor Edgard Santos.}

\author{Marcelo A. Moret}
\email{mamoret@gmail.com}
\affiliation{Modelagem Computacional, Universidade SENAI CIMATEC.}

\begin{abstract}
Quantum Machine Learning is a new computational tool that combines the quantum properties from quantum computing with the pattern recognition from machine learning. In this paper, we apply the Variational Quantum Classifier algorithm to the problem of pulsar classification of candidates from the High Time Resolution Universe 2 dataset. We use Qiskit Machine Learning circuits to compare the performance of the model using different feature selection methods, various number of features and training data size. Comparisons on the model from changing the data encoding and ansatz options are also reported. 


\end{abstract}

\maketitle

\section{Introduction}
Since the discovery of pulsars \cite{hewish1979observation} and their observation in a binary system \cite{Hulse1975}, many surveys have focused on the discovery of new candidates \cite{braun2015advancingtele, johnston2020thousandtele,bates2012highpulsarintro,stovall2014greenpulsarintro,deneva2009arecibopulsarintro,manchester2001parkespulsarintro}. Pulsars are rapidly rotating neutron stars, very characteristic for their precise cyclic signal bursts, useful for studying general relativity \cite{stairs2003testing, kramer2006tests}, gravitational waves \cite{jenet2004constraining} and navigation \cite{becker2018pulsar}.
The exponential rise in observable data made it impossible for candidates to be manually classified, thus generating the necessity of automatic processing of acquired data \cite{clancey1984classification,johnston1992highintro, braun2015advancing}. For this task, many machine learning applications are done using Neural Networks \cite{song2023effectivenessclassicoNN}, Support Vector Machines \cite{kannan2024classicalSVM}, and other classification algorithms \cite{9456250classical, wang2019hybridclassicogeral, chen2020researchclassicogeral,Thanu2023classicogeral, mcfarthing2024misto, Tariq2022,holewik2020imbalanced, mcfadden2017machine}. 
Quantum Machine Learning (QML) combines the potential quantum speedup, from quantum computation, with powerful pattern recognition, from machine learning \cite{Schuld2015,schuld2018supervisedQML, majid2025quantumQML}. QML is also used in the classification of pulsars \cite{tariq2022classicationquantum, Slabbert2024classquantum}. 

In this work, we use the High Time Resolution Universe 2 (HTRU-2) survey \cite{Lyon2016FiftyYO}. Studying the behavior and feature selection for this dataset.
A hybrid quantum-classical algorithm, not yet used for the classification task of pulsars, is the Variational Quantum Classifier (VQC). This algorithm is part of the large list of Variational Quantum Algorithms, and does classification through the use of a data encoding feature map combined with a quantum ansatz circuit. Those are then optimized with a classical sub-routine.
We present the underlying quantum circuits that form the VQC algorithm and show the configurations that obtain best accuracy. As the dataset is imbalanced, the Matthews Correlation Coefficient is also used for the comparison of results.

\section{Data and methods}
\subsection{Dataset}
Through the years of astrophysical observations and telescope evolutions, many pulsar surveys were conducted in an attempt to detect new candidates
\cite{braun2015advancingtele, johnston2020thousandtele,bates2012highpulsarintro,stovall2014greenpulsarintro,deneva2009arecibopulsarintro,manchester2001parkespulsarintro}.
However, feature choice and data volume became a problem for the classification of pulsar stars \cite{clancey1984classification,johnston1992highintro, braun2015advancing}.

In this study, we use the HRTU-2 dataset, which contains almost 18000 pulsar candidates with eight features, that is, distinct information about each candidate, in addition to one class for the labels \cite{Lyon2016FiftyYO}. The features are structured to mitigate problems that arise in other datasets.
Note that the class has values 0 for non-pulsars and 1 for the observed pulsars. For this reason, for performance evaluation, we consider pulsars as positive values and non-pulsars as negative.

Considering both integrated profile and Dispersion Measurement of the Signal to Noise Ratio (DM), the dataset brings the mean, standard deviation, skewness, and excess kurtosis, adding up to 8 features. These features were chosen by \cite{Lyon2016FiftyYO} to summarize data with the most distinct values, while avoiding the rise in the number of features. 
The first half of the features are related to
Profile, which refers to multiple values of the signal, forming an array, after being averaged in time and frequency.
The other features are related to the DM, a characteristic of error caused by interferences of signals in the interstellar medium. Table \ref{tab:abrevifeature} contains the perspective abbreviations of all the features.

\begin{table}[b]
    \caption{Abbreviations for the features in the HTRU-2 dataset.}
    \centering
    \begin{tabular}{|l|c|c|}
    \hline
         & \hspace*{8mm} Feature \hspace*{8mm} & \hspace*{3mm} Abbreviation \hspace*{3mm} \\
         \hline
         \hspace*{2mm}Profile & Mean & Prof$-\mu$ \\
          & Standard deviation & Prof$-\sigma$  \\
         &Kurtosis & Prof$-k$ \\
         & Skewness &Prof$-s$  \\
         \hline
         \hspace*{2mm}Dispersion & Mean &DM$-\mu$ \\
         \hspace*{1mm} measurement \hspace*{2mm} & Standard deviation & DM$-\sigma$ \\
         & Kurtosis & DM$-k$ \\
         & Skewness & DM$-s$  \\
         \hline
    \end{tabular}
    \label{tab:abrevifeature}
\end{table}

The observed data is distributed in a way that only 9.8\% of candidates are pulsar stars. The relatively low number of pulsars raise difficulties in identifying them \cite{lyon2016pulsars,guo2019pulsarSMOTE,azhari2020detectionSMOTE,beniwal2021detectionSMOTE}.

\begin{figure}[t]
        \centering
        \includegraphics[width=1\linewidth]{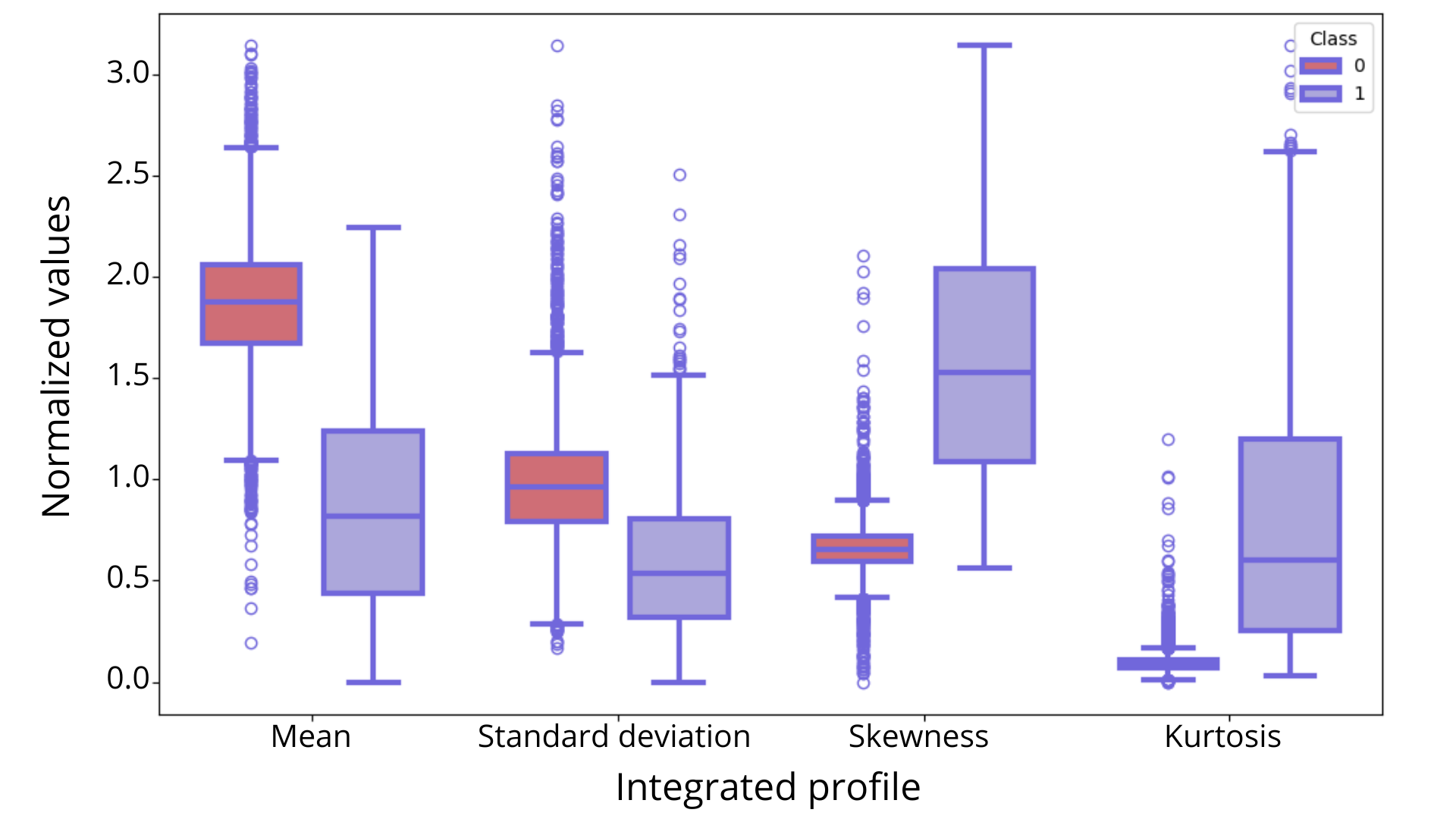}
        \caption{Box plot of the features related to the integrated profile. Red shows negative cases, non-pulsars, and purple positive pulsar candidates.}
        \label{fig:boxplot_profile}
\end{figure}

\begin{figure}[t]
    \centering
    \includegraphics[width=1\linewidth]{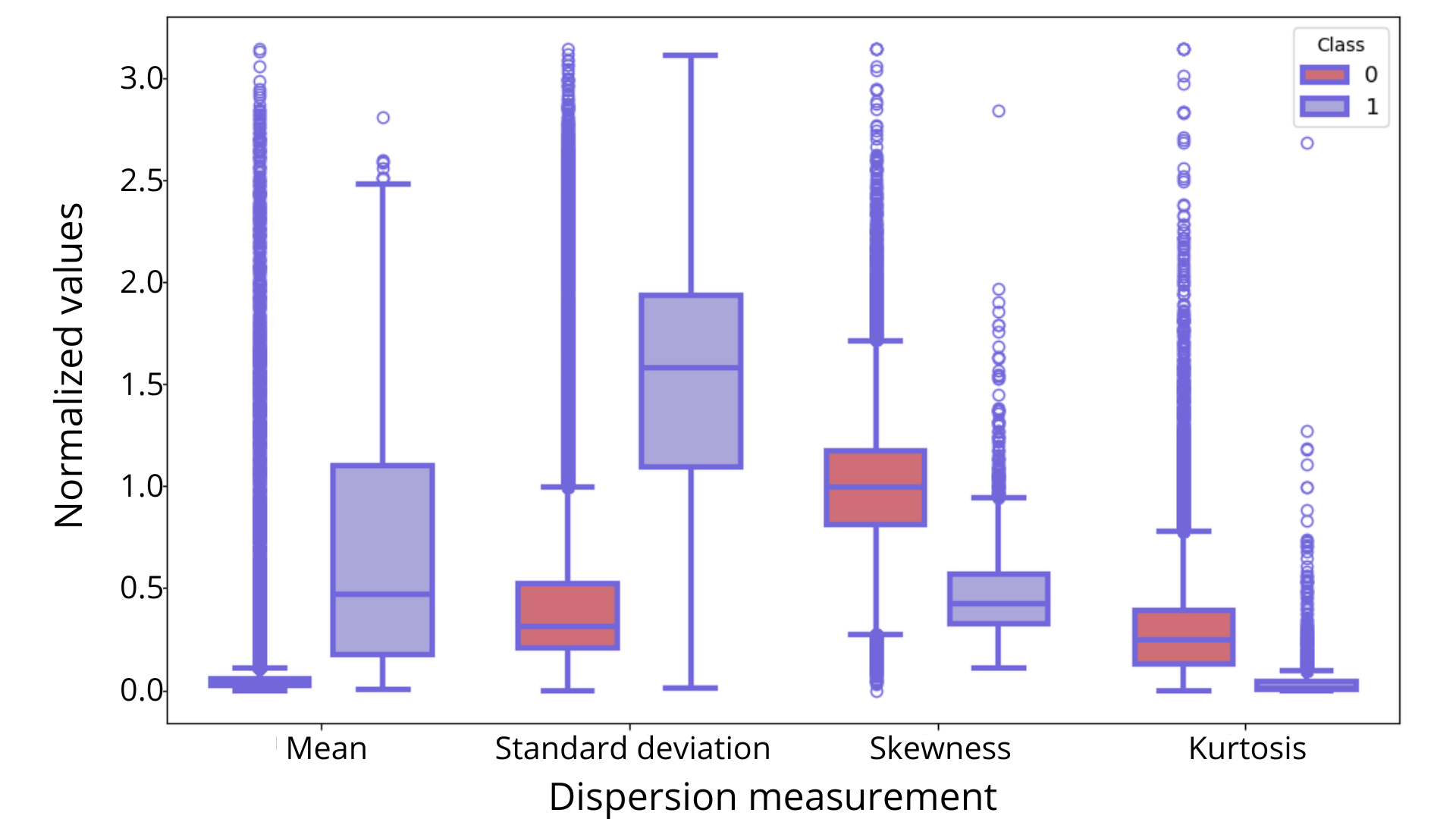}
    \caption{Box plot of features related to the DM. Red shows negative cases, non-pulsars, and purple positive pulsar candidates.}
\label{fig:boxplot_DM}
\end{figure}

The behavior of the dataset is expressed by the box plots in Fig. \ref{fig:boxplot_profile} and Fig. \ref{fig:boxplot_DM}. 
The box plot is an interesting statistical method to convey irregular data distribution
\cite{boxplot_krzywinski2014points}.
Notice that, for both the profile and DM features, data has a more well-balanced spread in the positive cases. Non-pulsars in this dataset tend to have a lot more outliers and smaller box plots for the asymmetrical quartiles.
    
\subsection{Methodology}
The first step in many machine learning procedures is data normalization.
It prevents the appearance of erroneous feature weights when dealing with different scales and feature ranges \cite{datanali2014data, datancabello2023impact, datanayak2014impact, datanahmed2022investigation}. 
As the dataset is filled with numerical data, further preprocessing, such as label encoding, will not be necessary.
The feature maps used are based on angle embedding, so the data was normalized using the \texttt{MinMaxScaler} from \texttt{sklearn}, in a range varying from $0$ to $\pi$.

The data is then divided into training and testing groups. The training portion of the dataset is first used to train the models. The machine gathers the information from the training dataset, acquiring knowledge about data interactions.
The newly trained model is then applied to the test portion of the dataset, returning a guess on the classification problem, which is then compared to the real values.

Performance comparison is done using a handful of binary classification metrics based on the four model output options: True Positive (TP) are cases correctly identified as a pulsar by the models; False Positives (FP) are classified as pulsars by the model, while they are not; False Negative (FN) are positive cases wrongly labeled as negative, that is, pulsar candidates which the model could not identify; and lastly True Negatives (TN) are candidates the model correctly classify as not being pulsars. Based on these four results, TP, TN, FP, and FN, we calculate accuracy, precision, recall, and F1-score \cite{Tharwat2018}. The ROC curve and its respective area are usually applied to measure the binary classification performance, but with imbalanced datasets, like the HRTU-2, it is not a good metric. The Matthews correlation coefficient (MCC) will be considered instead\cite{chicco2023matthews,chicco2020advantagesmatthews}.
A more thorough description of every metric used can be found in Appendix \ref{sec:ap_binaryclass}.

Different-sized samples were used to better determine when underfitting or overfitting occurs. In machine learning, overfitting happens if more data is used than necessary. In this situation, the model learns not only the behavior of the data, but also detects the noise, making the overall performance diminish. However, as the opposite can also happen, blindly reducing the data used in training is not a good method.
If the amount of data is insufficient, it prevents the model from learning the data behavior in any manner \cite{bashir2020information,montesinos2022overfitting}.

All the algorithms were implemented using Qiskit tools. The VQC from \texttt{qiskit-quantum-machine-learning.algorithms} was first applied using only data normalization and then compared to tests using feature selection. For a more complete, wider overview of the performance of the algorithm, we also repeated its implementation for a set of combinations on the number of qubits used,\footnote{The number of qubits in the system is directly related to the number of features. For the feature maps used, each qubit represents one feature.} the portion of the dataset employed in training and the different configurations of data encoding and ansatz.

\subsection{Feature selection}  \label{sec:featureselection}
Feature selection is a machine learning procedure that reduces the number of features. This reduction aims to decrease the computational resources used and training time, while maintaining a good performance
\cite{kohavi1997wrappersfeatureselec, guyon2003introductionfeatureselec, mucke2023featureselec, chandrashekar2014surveyfeatureselec, zhao2010advancing, langley1994selection}. 
As quantum computing is in the Noisy Intermediate-Scale Quantum (NISQ) era, circuits are noise-sensitive. Feature selection becomes important as a tool to minimize the number of qubits used, consequently lowering the error caused by noise\cite{preskill2018quantumnisq, lau2022nisq}.

\begin{figure}[t]
    \centering
    \includegraphics[width=\linewidth]{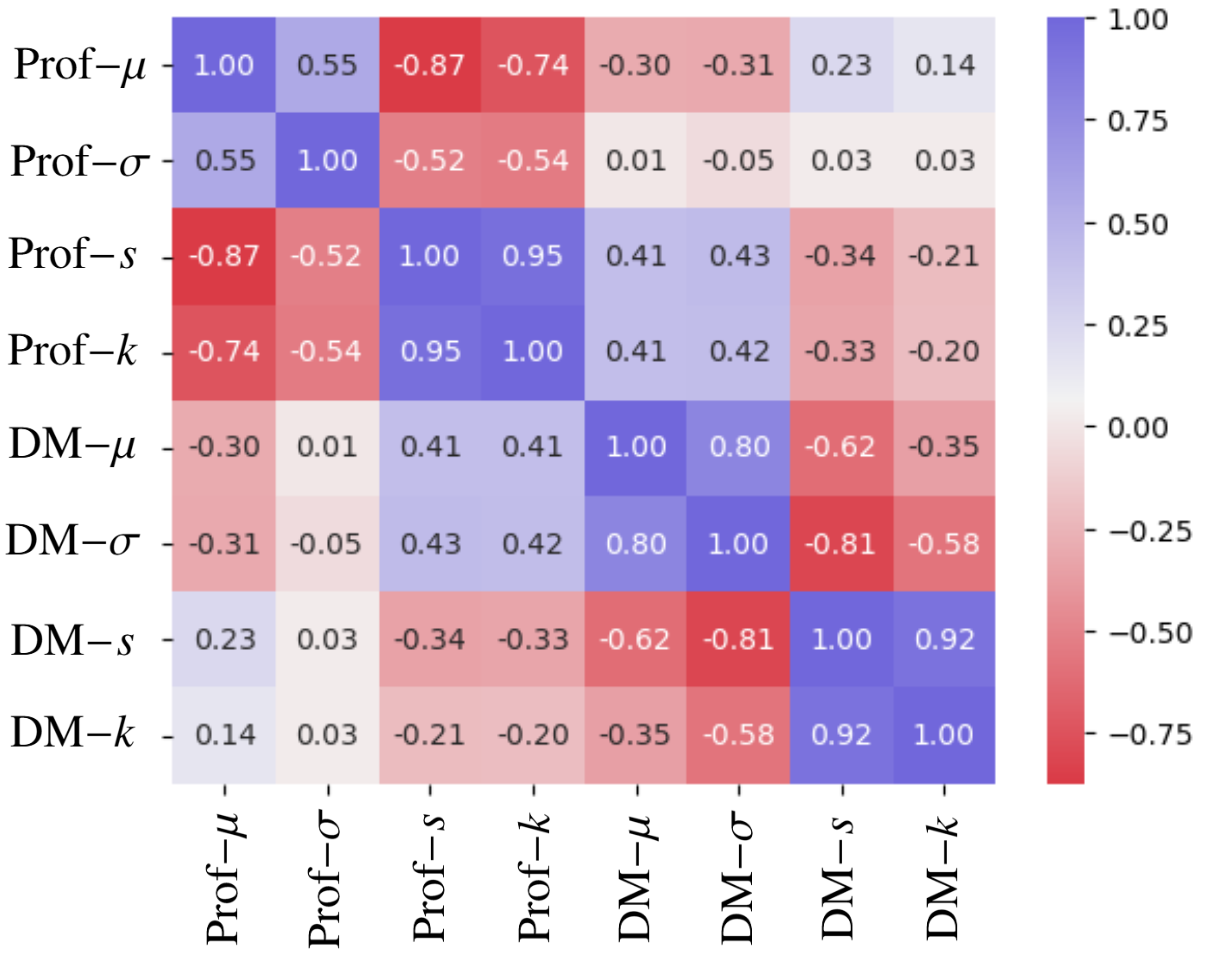}
    \caption{Pearson correlation matrix of all features in the dataset.}
    \label{fig:correlation}
\end{figure}

The correlation matrix is a common representation of how features associate with each other. When two or more features have a high absolute value of correlation, they carry repetitive information and, therefore, diminish the model performance. Positive correlation shows that variables change in the same direction, rising or lowering together, while negative correlation displays the opposite: the increase of one feature leads to the decrease of the other.
The study of correlation is a common practice in feature selection, as redundant features cause overfitting
\cite{hall1999correlation, kaur2021feature, cai2018feature} .
The Pearson correlation of all the features can be found in Fig. \ref{fig:correlation} \cite{cohen2009pearson}.

\begin{figure}[t]
    \centering
    \includegraphics[width=0.38\linewidth]{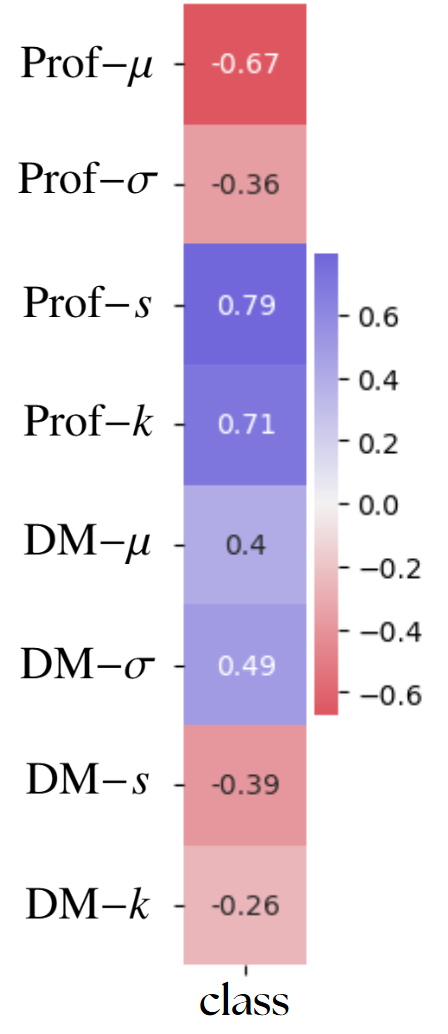}
    \caption{Pearson correlation between all the features and the target class.}
    \label{fig:corr_class}
\end{figure}

Two feature selection methods are used and compared. The first one, later referred to as FS1, is obtained 
through the use of \texttt{selectkbest} from \texttt{sklearn.preprocessing} with the \texttt{f\_classif} function. The second method, FS2, is defined by the highest to lowest modulus of correlation for each feature with the class indicator, which is shown in Fig. \ref{fig:corr_class}.
Every model trained using fewer than eight features
followed the sequence expressed in Table \ref{tab:features}, which also contains the original sequence on the HTRU-2 dataset.

\begin{table}[t]
\centering
    \caption{Feature order used for feature selection with \textit{SelectKBest} and correlation to class.}
    \begin{tabular}{|c |c |c|}
    \hline
         \hspace*{2mm}Original order \hspace*{2mm} & \hspace*{5mm}FS1\hspace*{6mm} & \hspace*{5mm}FS2 \hspace*{6mm}\\
         \hline
         Prof.$_\mu$ & Prof.$_\mu$ & Prof.$_s$ \\
         Prof.$_\sigma$ & Prof.$_s$  & Prof.$_k$ \\
         Prof.$_s$ & Prof.$_k$ & Prof. $_\mu$  \\
         Prof.$_k$ & DM $_\mu$ &  DM $_\sigma$ \\
         DM $_\mu$ & DM $_\sigma$ & DM $_\mu$ \\
         DM $_\sigma$ & DM $_s$ & DM $_s$ \\
         DM $_s$ & DM $_k$ &  Prof. $_\sigma$\\
         DM $_k$ & Prof.$_\sigma$ & DM.$_k$ \\
        \hline
    \end{tabular}
    
    \label{tab:features}
\end{table}

\section{Variational Quantum Classifier}
The Variational Quantum Classifier is a hybrid quantum-classical routine, belonging to the many Variational Quantum Algorithms that exist \cite{mcclean2016variationaltheory, cerezo2021variational,adebayo2023variational}. The first and most famous VQA is the Variational Quantum Eigensolver (VQE), which estimate the ground state of a Hamiltonian
\cite{fedorov2022vqe}.
The VQC is a relatively newer algorithm \cite{saxena2022performance,maheshwari2021variational}, but it already contains various applications.
Some uses of the VQC include the energy industry \cite{VQCapphangun2024quantum}, climate action \cite{VQCappmunasinghe2024assessment}, agriculture \cite{mukhamedieva2024application}, healthcare \cite{khan2025diabetes}, and astrophysics
\cite{VQCastrobelis2021higgs, VQCastroregadio2025exoplanet, VQCastrobhavsar2023classification}.

The VQC is a circuit that uses a Feature map into encode data to the quantum space and an ansatz. 
Classical routines are used to optimize and update the parameters from the ansatz.
The circuit available in \texttt{qiskit\_machine\_learning.algorithms} 
gives the opportunity to define an associated loss function and optimizer, which are the classical sub-routines. In this work, we limited our tests to the default values of the loss function, as the cross entropy, and the SLSQP, as the optimizer.

\subsection{Feature maps} \label{sec:featuremap}
The Qiskit Machine Learning package contains a wide variety of ready-to-use circuits that can be applied as both feature maps or ansatz for the VQC. For more information and user guide refer to \cite{IBMQuantumDocumentation}. 

The first group of circuits used follows angle embedding properties. Based on the Pauli rotation matrices \eqref{eq:paulimatrices}, both the Pauli Feature Map and ZZ Feature Map apply rotations for data embedding \cite{suzuki2020analysis}. The Pauli Feature Map is a highly adjustable data encoding circuit that maps classical information into a quantum space. Its circuit is based on the application of Hadamard gates, followed by a Phase gate dependent on the qubit. The next step is the implementation of entanglement circuits, which connect qubits with one another. The entanglement circuit is followed by phase gates, which depend on the connected information of both entangled qubits. The rotation applied can be chosen as X, Y, or Z for the first order, XX, YY, or ZZ for the second order, or any combination of these directions. More about the entanglement routine can be found in \ref{sec:apentanglement}.

\begin{equation}
\sigma_X = \begin{pmatrix}
    0 & 1 \\ 1 & 0
\end{pmatrix},\quad
\sigma_Y = \begin{pmatrix}
    0 & -i \\ i & 0
\end{pmatrix},\quad
    \sigma_Z = \begin{pmatrix}
        1 & 0 \\ 0 & -1
    \end{pmatrix}.\label{eq:paulimatrices}
\end{equation}

\begin{figure}[t]
    \centering
    \includegraphics[width=\linewidth]{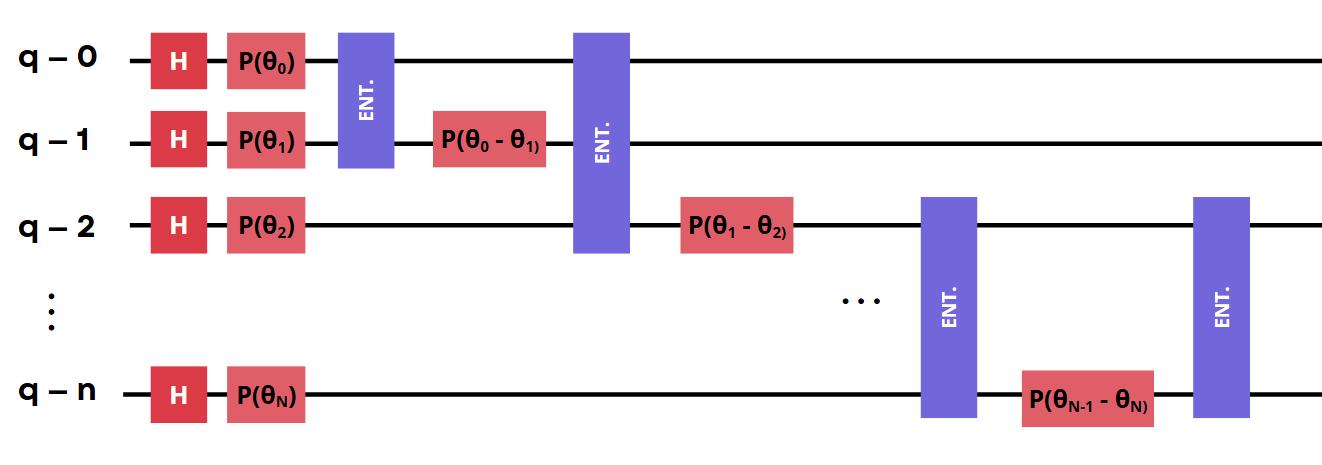}
    \caption{Quantum circuit for the angular data encoding sub-routines, ZZ and Pauli feature maps, in N qubits.}
    \label{fig:circuit-featureZZPauli}
\end{figure}

The ZZ Feature Map is a second-order Pauli circuit, applying specifically ZZ rotations. Its circuit is similarly built with a series of Hadamard and Phase gates, followed by entanglement and rotation gates. Fig. \ref{fig:circuit-featureZZPauli} shows the scheme for an angle embedding circuit in N qubits. This circuit is the same for the default Pauli and ZZ feature maps used in this paper.

\subsection{Ansatz}

The ansatze are the initial circuits, whose parameters will be updated in a way to approximate the model to the desired loss function. The feature map circuits can be used as an ansatz, but the other options available perform better.

\begin{figure}[t]
    \centering
    \includegraphics[width=0.4\linewidth]{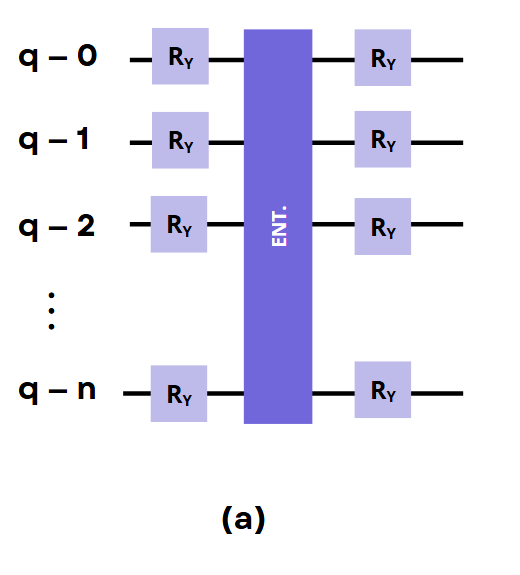}
    \includegraphics[width=0.57\linewidth]{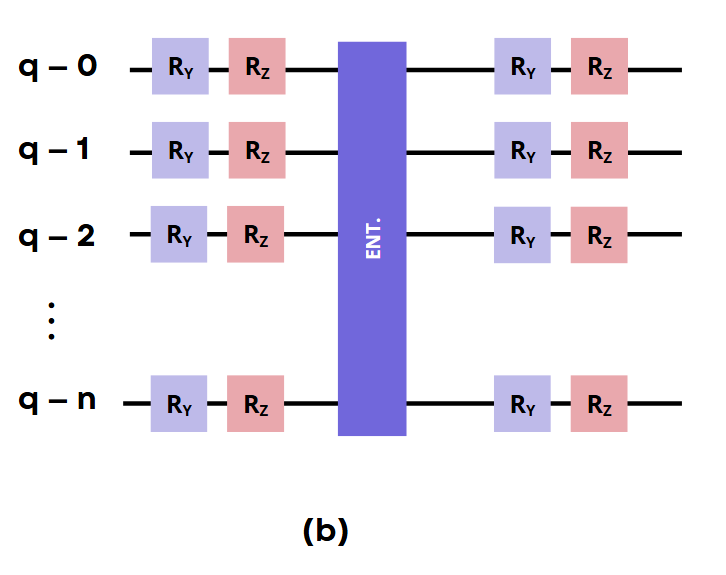}
    \caption{Quantum circuits for ansatz sub-routine applied to a N qubit system. On the left,
    (a) is the Real Amplitudes ansatz, with Y-rotation gates before and after an entanglement circuit and, on the right, (b) is the Efficient SU2 formed by two sets of rotations before and after the entanglement circuit.}
    \label{fig:ansatz_circ}
\end{figure}

The Real Amplitudes circuit is the first ansatz used. Its prepared states have only the real amplitudes, while the complex portion is always zero. The circuit in Fig. \ref{fig:ansatz_circ} (a) shows the rotations and entanglement sequence that form the Real Amplitudes routine.

The Efficient Special Unitary 2 or Efficient SU2 is a Two-local based circuit applied usually to variational and classification algorithms. Similar to the Real Amplitudes circuit, for the Efficient SU2 in Fig. \ref{fig:ansatz_circ} (b) a set of entanglement is applied between rotations, but this time there are Y and Z rotations.

Both sub-routines in Fig. \ref{fig:ansatz_circ} show one repetition of the equivalent ansatz, but the same set of gates can be applied consecutively to form a bulkier circuit. In all experiments done, the repetition was set to two, for both ansatz and feature maps.

\section{Results}
\begin{figure}[t]
    \centering
    \includegraphics[width=\linewidth]{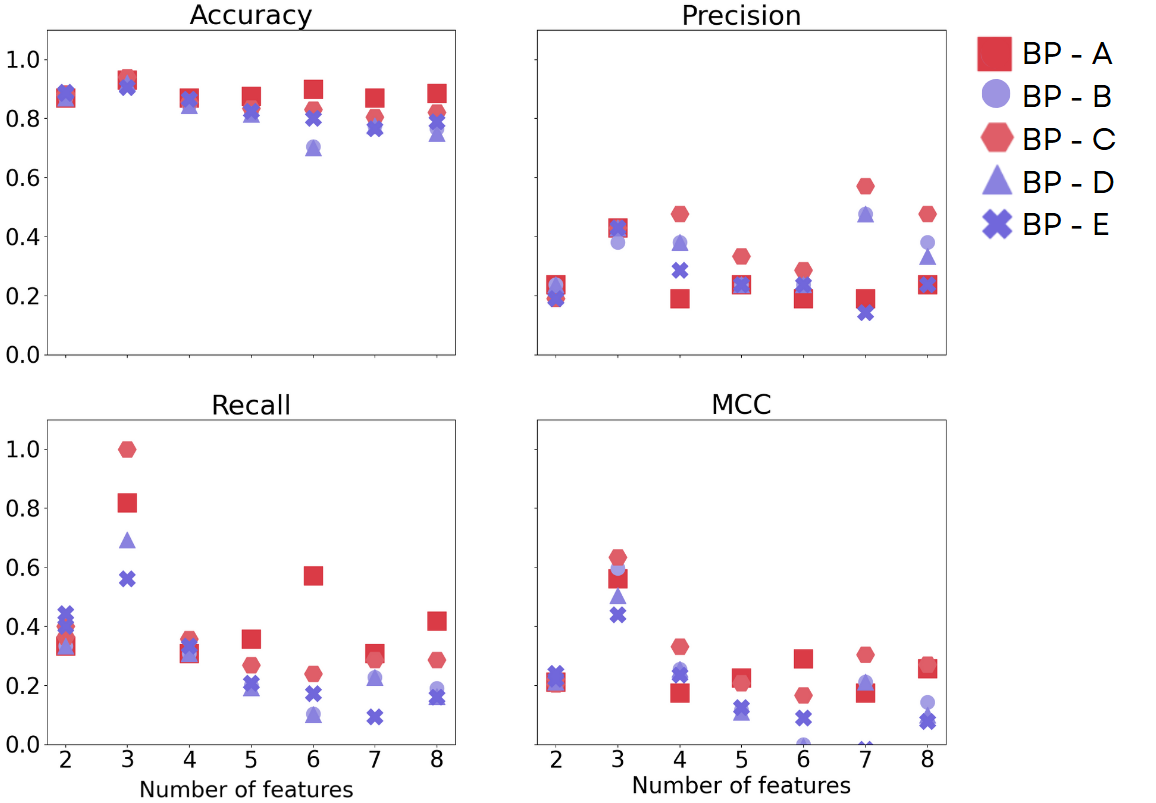}
    \caption{Best-performing VQC configurations for a 300 data sized training batch. The complete description of the labels can be found on Table \ref{tab:compara_geral}.}
    \label{fig:Res_geral}
\end{figure}

The first batch of tests done with the VQC used the FS1 as described in \ref{sec:featureselection}. We varied the number of candidates used for testing with 180 and 300, as well as the number of features, ranging from 2 to 8. Some of the best-performing (BP) models, based on accuracy for all numbers of features, are explicit in Fig. \ref{fig:Res_geral}. Note that some combinations with higher accuracy for one set of qubits performed poorly for other quantities, and thus are not present in this first list, but are included in \ref{tab:compara_geral_total}.
The variation in accuracy, precision, recall, and MCC can be analyzed as the number of features rises. For a full description of the labels used, please refer to Table \ref{tab:compara_geral}. It is interesting to note that, even though accuracy does not present significant changes through the variation in number of qubits, the recall and MCC have higher values for 3 features.

\begin{figure}[t]
    \centering
    \includegraphics[width=1\linewidth]{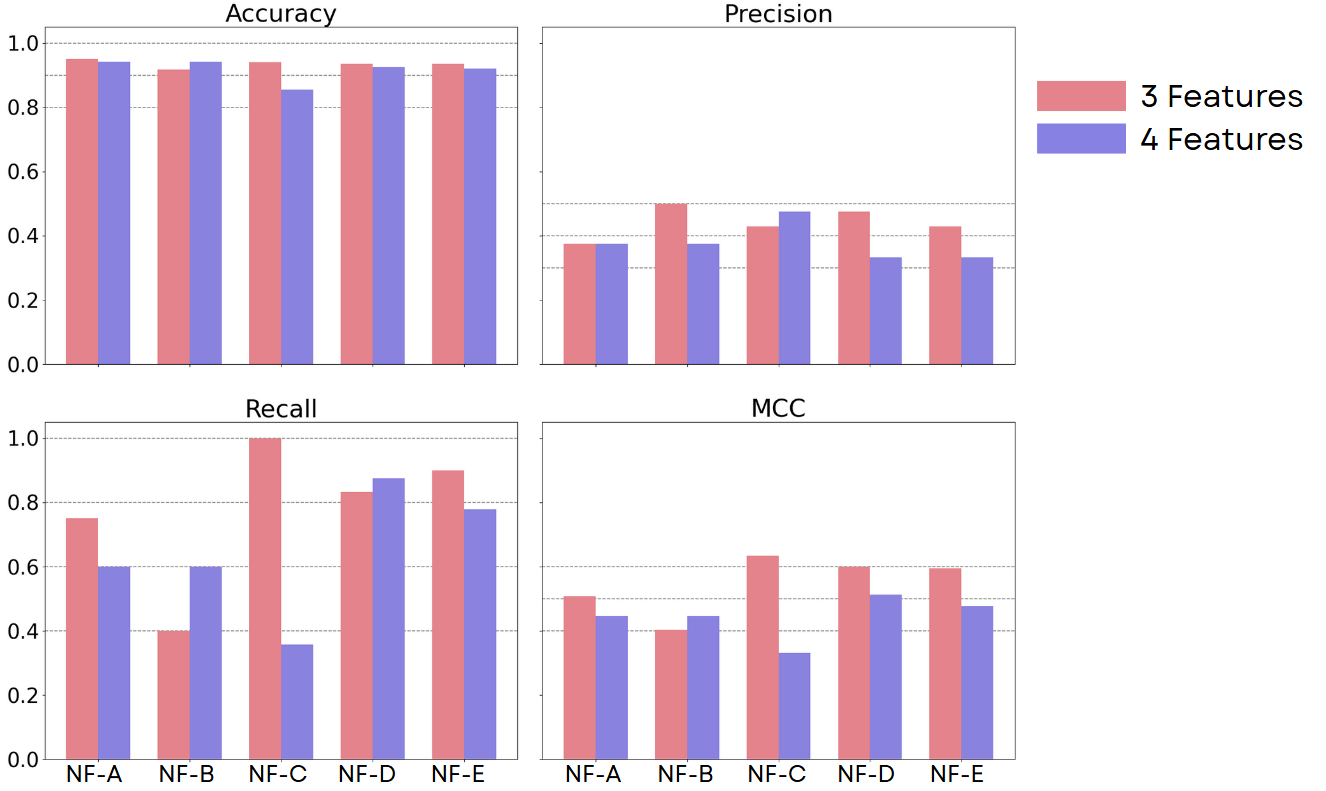}
    \caption{Comparison between three and four features fixed to the FS1 feature selection. The best-performing VQC models using for three qubits is representend in pink, while purple shows the results for four features. A full description of the labels, containing size of the training batch, feature map and ansatz can be found on Table \ref{tab:compara_numfeat}}
    \label{fig:Res_numfeat}
\end{figure}

As the best-performing models were obtained with 3 and 4 qubits, we used this setting for further tests.
The first step is the comparison between these configurations. Fig. \ref{fig:Res_numfeat} contains the five best overall performances for the VQC with 3 and 4 features. Even though there was variation in every aspect (data size, feature map, ansatz and entanglement), all the best-performing circuits were built with the Efficient SU2 ansatz, while Real amplitudes made no appearances. Full description of the labels for Fig. \ref{fig:Res_numfeat} is present in Table \ref{tab:compara_numfeat}. The lower number of qubits also corresponds to lower training time, which is useful for running multiple combinations of all the other characteristics.

\begin{table}[t]
    \centering
    \caption{Labels for the best-performing overall combinations, using training data size of 300 and the FS1 feature selection. }
    \begin{tabular}{|c|c|c|c|}
    \hline
        Label & Feature Map & Ansatz & Entanglement \\
        \hline
        BP-A & ZZ & Real amplitudes & Linear \\
        BP-B & ZZ  & Real amplitudes & Linear \\
        BP-C & ZZ  & Efficient SU2 & Circular \\
        BP-D & ZZ  & Real amplitudes & Circular \\
        BP-E & Pauli  & Efficient SU2 & Circular\\
        \hline
    \end{tabular}    
    \label{tab:compara_geral}
\end{table}

\begin{table}[t]
    \centering
    \caption{Labels for the best-performing VQC models in common for 3 and 4 features, using the FS1 feature selection. }
    \begin{tabular}{|c|c|c|c|c|}
    \hline
         Label & Data size & Feature Map & Ansatz & Entanglement\\
         \hline
         NF-A & 180 & ZZ & Efficient SU2 & Circular \\
         NF-B & 180 & Pauli & Efficient SU2 & Circular \\
         NF-C & 300 & ZZ & Efficient SU2 & Circular \\
         NF-D & 300 & ZZ & Efficient SU2 & Full \\
         NF-E & 300 & Pauli & Efficient SU2 & Full \\
         \hline
    \end{tabular}
    
    \label{tab:compara_numfeat}
\end{table}

\begin{figure}[t]
    \centering
    \includegraphics[width=0.97\linewidth]{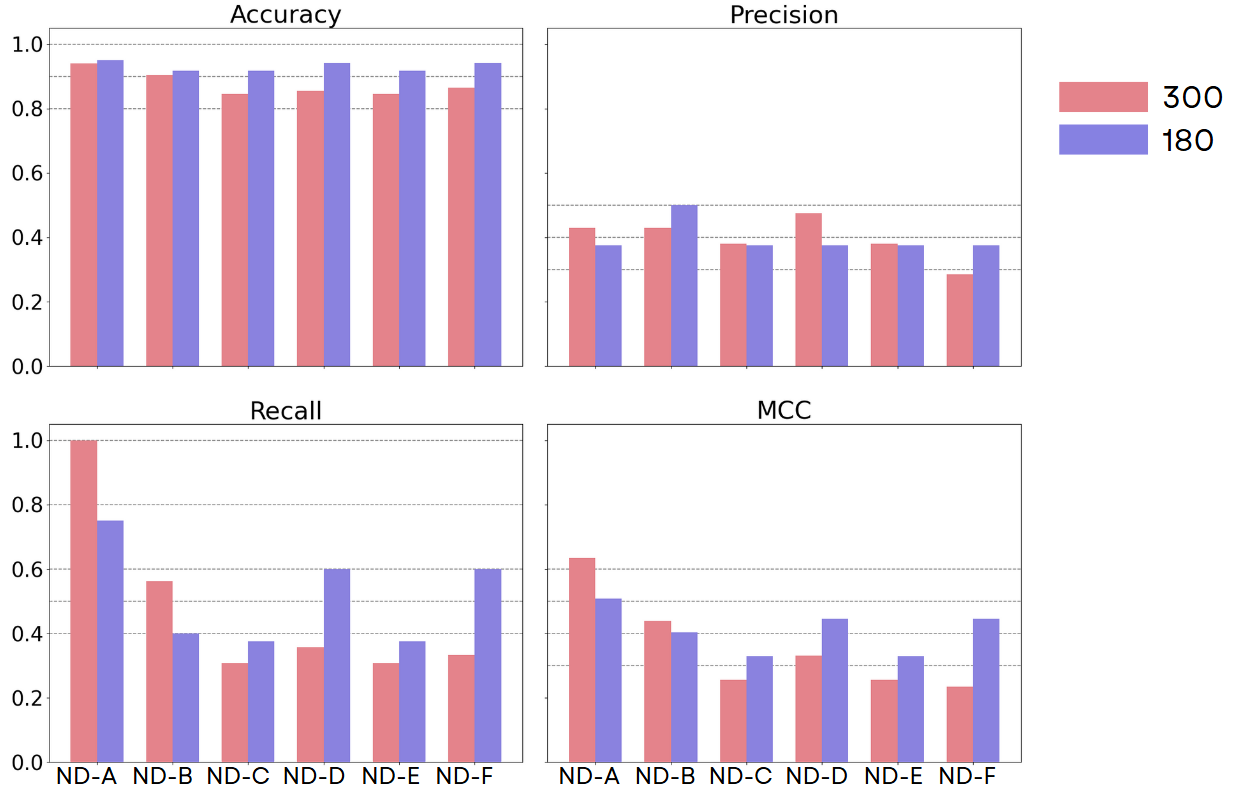}
    \caption{Comparison between the best-performing models for different data training size. The tests were done
    using the FS1 feature selection method. Labels containing number of features, feature map and ansatz can be found on Table \ref{tab:compara_numdados}.}
    \label{fig:Res_numdados}
\end{figure}

The effects of data size in training are explicit in Fig. \ref{fig:Res_numdados}, which brings results for the 180 and 300 batch sizes. The corresponding labels are in Table \ref{tab:compara_numdados}.
In all cases, the smaller-sized batch had better accuracy, but varied in all other metrics.

\begin{table}[b]
    \centering
    \caption{Best-performing models for 180 and 300 sized training batch using the FS1 feature selection.}
    \begin{tabular}{|c|c|c|c|c|}
    \hline
         Label & Features & Feature Map & Ansatz & Entanglement\\
         \hline
         ND-A & 3 & ZZ & Efficient SU2 & Circular \\
         ND-B & 3 & Pauli & Efficient SU2 & Circular \\
         ND-C & 4 & ZZ & Real amplitudes & Circular \\
         ND-D & 4 & ZZ & Efficient SU2 & Circular \\
         ND-E & 4 & Pauli & Real amplitudes & Circular \\
         ND-F & 4 & Pauli & Efficient SU2 & Circular \\
         \hline
    \end{tabular}
    
    \label{tab:compara_numdados}
\end{table}

Another important comparison was between the different feature selection options. Fixing the training size to 300, the metrics for the best-performing models are found in Fig. \ref{tab:compara_featselect}. Accuracy is similar to both configurations, but while FS2 has a slightly better precision, FS1 has better recall and MCC.

\begin{figure}[t]
    \centering
    \includegraphics[width=0.96\linewidth]{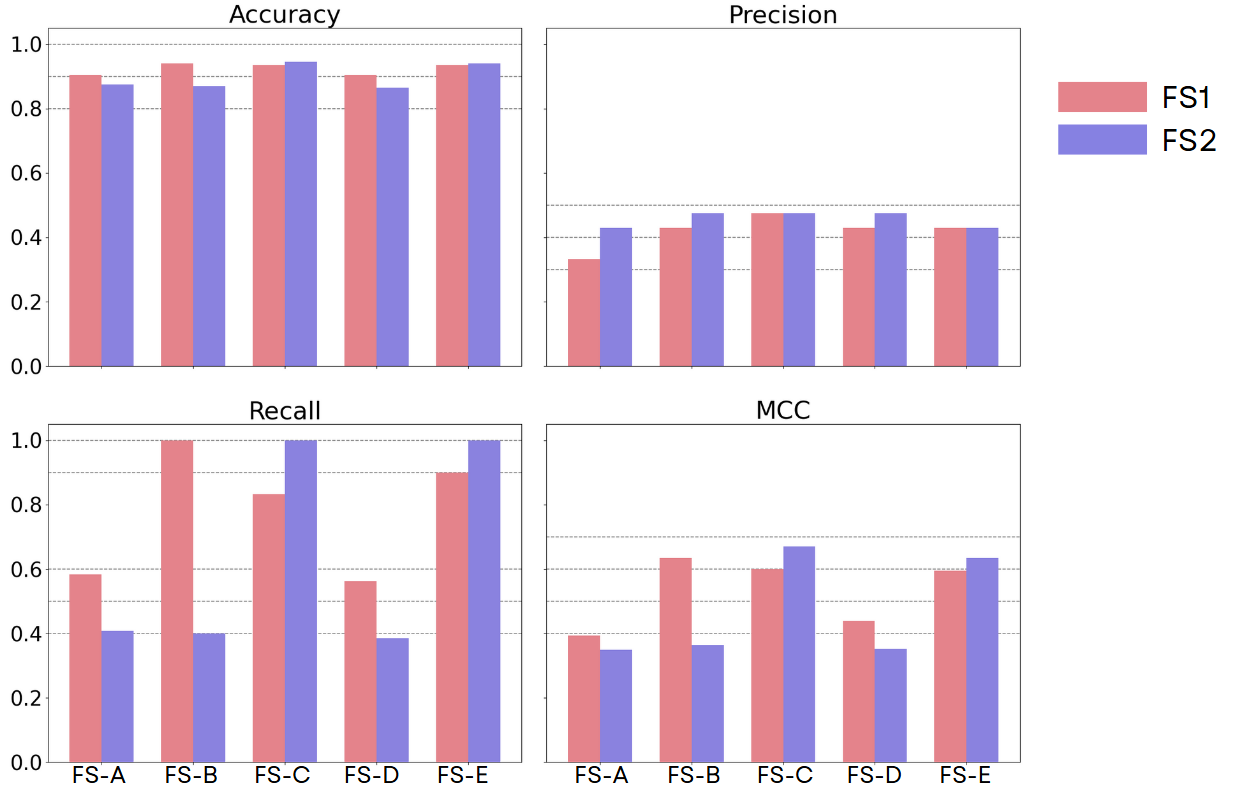}
    \caption{Comparison of the est performing VQC models for the feature selection options FS1 and FS2 using a fixed 300 size batch for training. All the labels with number of features, feature map and ansatz are explained in Table \ref{tab:compara_featselect})}
    \label{fig:Res_featselec}
\end{figure}

\begin{table}[b]
    \centering
    \caption{Labels for the best-performing VQC models in common for FS1 and FS2 with a 300 sized training batch.}
    \begin{tabular}{|c|c|c|c|c|}
    \hline
         Label & Features & Feature Map & Ansatz & Entanglement\\
         \hline
         FS-A & 3 & ZZ & Efficient SU2 & Linear \\
         FS-B & 3 & ZZ & Efficient SU2 & Circular \\
         FS-C & 3 & ZZ & Efficient SU2 & Full \\
         FS-D & 3 & Pauli & Efficient SU2 & Linear \\
         FS-E & 3 & Pauli & Efficient SU2 & Full \\
         \hline
    \end{tabular}
    
    \label{tab:compara_featselect}
\end{table}

As both the feature map and ansatz routines are quantum circuits, they are interchangeable. Tests done using the ZZ Feature map and Pauli feature map as ansatz had an interesting result: the overall performance was bad, but the precision was higher than average. The repetition of ZZ as a feature map and ansatz got 0.625 precision when applied to all eight features. The correspondent accuracy was only 0.425, while the MCC was 0.018. Using FS2, all the tests combining ZZ and Pauli feature map as ansatze got over 0.500 precision when using only two qubits, while the accuracies ranged from 0.860 to 0.890. Considering the description of FS2 in Table \ref{tab:features}, this means that profile skewness and kurtosis were the only info necessary to achieve high precision for repeating the ansatz. These results carry the information that using different circuits lets the model expand on the given data, getting better overall results, while using the same circuits identifies more positive candidates.

Lastly, Table \ref{tab:compara_geral_total} has the numerical values of the best performance for every combination discussed. Note that some of these are present in the comparison figures and table. Some others, however, have not been included yet, as the counterpart for any comparison was not high enough, and got filtered out, as a way to not introduce bias to the displayed results.

\begin{table*}[!t]
    \centering
    \caption{Best-performing models based on accuracy and MCC.}
    \begin{tabular}{|c|c|c|c|c|c||c|c|c|c|c|}
    \hline
         Features & Data size & Feature selection & Feature Map & Ansatz & Entanglement & Accuracy & Precision & Recall & F1-score & MCC \\
         \hline
          3 & 180 & FS1 & ZZ & EfficientSU2 & circular & 0.950 & 0.375 & 0.750 & 0.500 & 0.509 \\
          3 & 300 & FS2 & ZZ & EfficientSU2 & full & 0.945 & 0.476 & 1.000 & 0.645 & 0.670 \\
        3 & 300 & FS2 & Pauli & EfficientSU2 & full & 0.940 & 0.429 & 1.000 & 0.600 & 0.634 \\
        3 & 300 & FS1 & ZZ & EfficientSU2 & circular & 0.940 & 0.429 & 1.000 & 0.600 & 0.634 \\
        
        3 & 300 & FS1 & ZZ & EfficientSU2 & full & 0.935 & 0.476 & 0.833 & 0.606 & 0.600 \\
        3 & 300 & FS2 & ZZ & RealAmplitudes & full & 0.935 & 0.476 & 0.833 & 0.606 & 0.600 \\
        3 & 300 & FS1 & Pauli & RealAmplitudes & circular & 0.935 & 0.381 & 1.000 & 0.552 & 0.596 \\
        2 & 300 & FS2 & Pauli & EfficientSU2 & full & 0.935 & 0.381 & 1.000 & 0.552 & 0.596 \\
        3 & 300 & FS1 & Pauli & EfficientSU2 & full & 0.935 & 0.429 & 0.900 & 0.581 & 0.595 \\
  
        3 & 300 & FS1 & ZZ & RealAmplitudes & linear & 0.930 & 0.429 & 0.818 & 0.563 & 0.561 \\
        3 & 300 & FS2 & Pauli & RealAmplitudes & full & 0.930 & 0.429 & 0.818 & 0.563 & 0.561 \\
        2 & 300 & FS2 & ZZ & EfficientSU2 & linear & 0.930 & 0.333 & 1.000 & 0.500 & 0.556 \\
        2 & 300 & FS2 & ZZ & EfficientSU2 & circular & 0.930 & 0.381 & 0.889 & 0.533 & 0.555 \\
        2 & 300 & FS2 & Pauli & EfficientSU2 & linear & 0.930 & 0.381 & 0.889 & 0.533 & 0.555 \\
        4 & 300 & FS1 & ZZ & EfficientSU2 & full & 0.925 & 0.333 & 0.875 & 0.482 & 0.513 \\
        3 & 300 & FS1 & ZZ & RealAmplitudes & circular & 0.920 & 0.429 & 0.692 & 0.530 & 0.505 \\

          \hline
    \end{tabular}
    
    \label{tab:compara_geral_total}
    \vspace*{4pt}
\hrulefill
\vspace*{4pt}
\end{table*}

\section{Conclusion}
The VQC circuit, from the Qiskit library, is easy to use and presents a good performance on the classification problem for pulsars, achieving accuracies as high as 0.950. Using the correct combination of data encoding, ansatz, and feature selection is very important, as the same model can have a 0.050 disparity in accuracy with the worst configuration. The \texttt{selectkbest} feature selection had the best scores with 3 and 4 features, while applying ansatze that differ from the data encoding circuit.

The use of quantum machine learning is fairly recent, but already promising. The power of data treatment with quantum computers may help bring new advances, not only in classification problems, but to astrophysics in general.
An expansion of this work is the analysis of both classical sub-routines, loss function, and optimizer, and how they impact the performance of VQC. This study can also be extended by implementing the circuits on real quantum computers or noisy simulators. Future tests on fault-tolerant quantum computers could bring even better performances and be officially implemented as tools for data processing in astrophysics.

\appendix
\section{Binary classification} \label{sec:ap_binaryclass}

The performance metrics are calculated using information from the four main outputs: TP, FP, FN, and TN. 
Most acquired values vary between 0 and 1, where higher values signify better performance. The MCC is the only exception, ranging from -1 to 1.

The first metric used is the accuracy. It is a general performance metric that shows the overall correctness of a model. Ranging from 0 to 1, where 1 is a perfect classifier, accuracy is given by 
\begin{equation}
 Acc = \dfrac{TP + TN}{TP + FP + TN + FN} .
\end{equation}
Precision dictates how well the model predicts positive cases, compared to false positives. Its formula is 
\begin{equation}
 Prec = \dfrac{TP}{TP + FP}   .
\end{equation}
For low values of precision, we could identify many of the pulsars, but also waste resources on false candidates. The next metric is recall, which measures how many of the positive cases were actually detected. Calculated with 
\begin{equation}
 \dfrac{TP}{TP + FN},   
\end{equation}
recall can be prioritized over precision if identifying every pulsar star is more important than the amount of resources used. For imbalanced data, precision and recall may have very distinct values; for that reason F1-score is also analyzed. Given by the average of recall and precision, 
\begin{equation}
    \dfrac{2 \times Precision \times Recall}{Precision + Recall}, 
\end{equation}this metric balances the desire for extracting all positive values and not wasting resources.
The last metric, MCC is a coefficient designed for binary classification, specially when applied to imbalanced data. It takes values from +1, perfect classifier with total agreement, to -1 total disagreement. It is calculated with \eqref{eq:MCC}.

\begin{equation}
    MCC = \dfrac{TP \times TN \textnormal{ -- } FP \times FN}{\sqrt{(TP + FP) (TP + FN) (TN + FP)  (TN +FN)}}
    \label{eq:MCC}
\end{equation}

\section{Entanglement} \label{sec:apentanglement}
Entanglement is a purely quantum property used in several quantum computing algorithms. In this work, it appears as a sub-routine to the feature maps that encode data. The entanglement options vary in how they connect qubits - or features - with each other. Three different entanglement circuits were tested: linear, circular, and full. These connections happen when the Controlled NOT (CNOT) gate is used. CNOT acts on two qubits, applying an X gate in the target qubit if the control qubit has value $\ket{1}$.

\begin{figure}[t]
    \centering
    \includegraphics[width=0.43\linewidth]{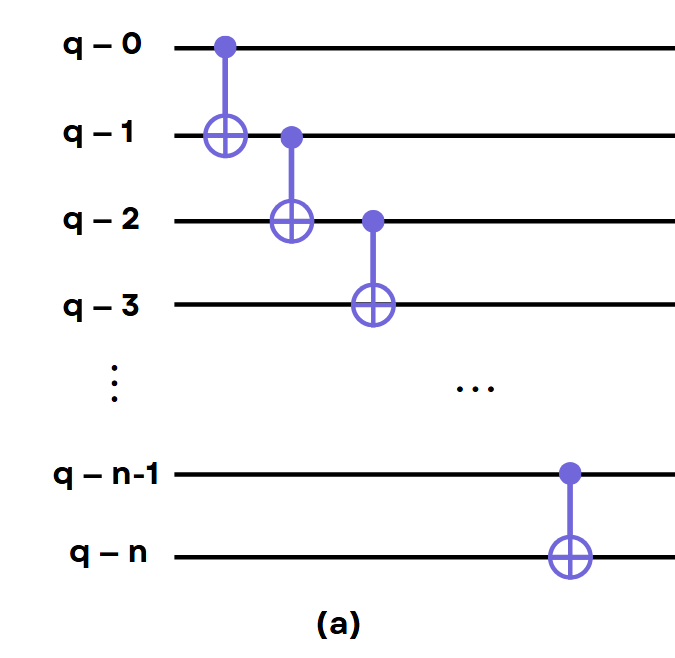}
    \includegraphics[width=0.53\linewidth]{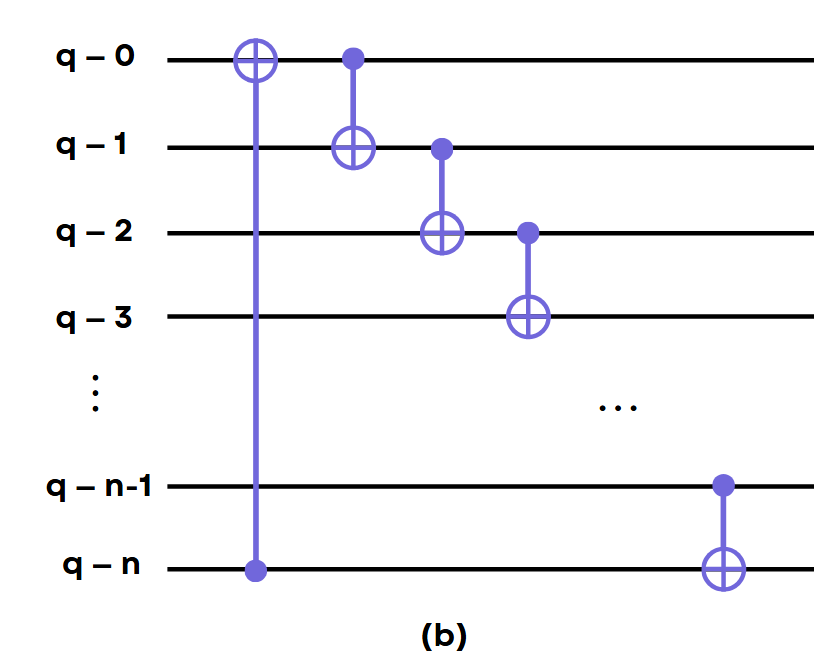}
    \caption{CNOT-gate dispositions in a N qubits quantum circuit for: (a) linear entanglement and (b) circular entanglement.}
    \label{fig:esquema-linear_ent}
\end{figure}

Linear entanglement, as the name suggests, connects each qubit linearly with the subsequent ones. This behavior is explicit in Fig. \ref{fig:esquema-linear_ent} (a). Circular entanglement acts in a similar way, but it also adds a layer of entanglement for the first and last qubits, connecting them as shown in Fig. \ref{fig:esquema-linear_ent} (b).

\begin{figure}[t]
    \centering
    \includegraphics[width=0.9\linewidth]{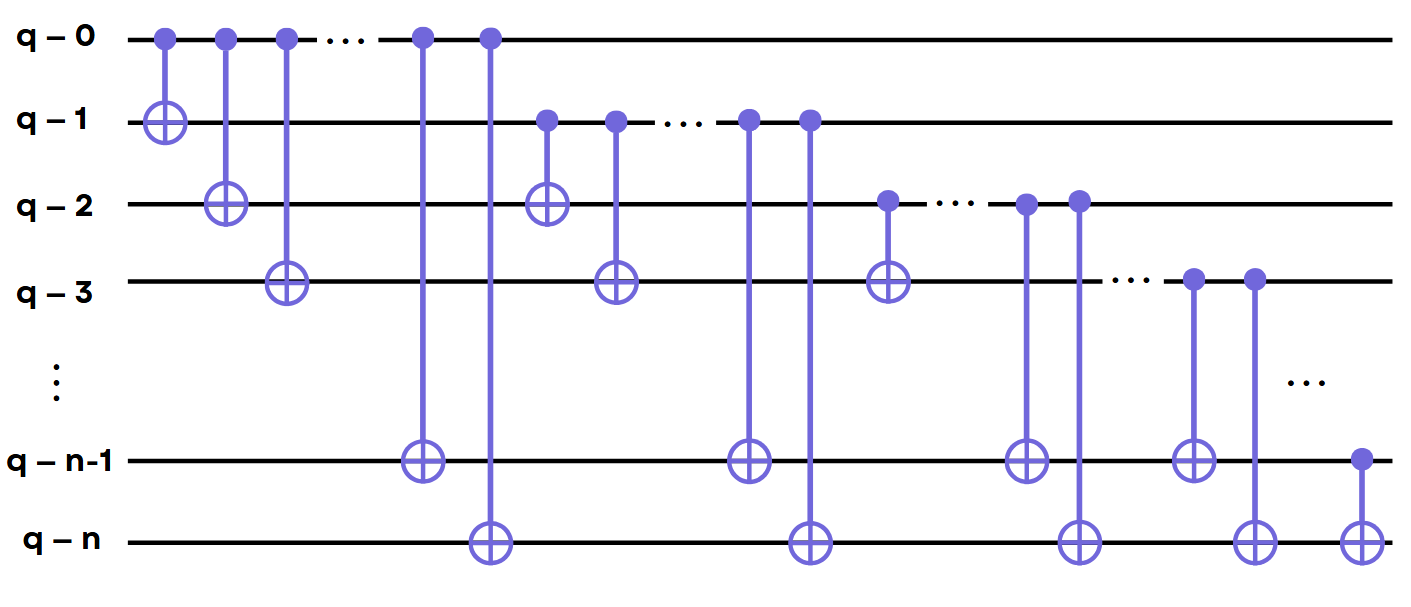}
    \caption{CNOT-gate disposition for the full entanglement circuit with N qubits.}
    \label{fig:esquema-full_ent}
\end{figure}

Full entanglement, on the other hand, provides a stronger entanglement, connecting every qubit with each other; this is shown in Fig. \ref{fig:esquema-full_ent}.

\begin{figure}[t]
    \centering
    \includegraphics[width=\linewidth]{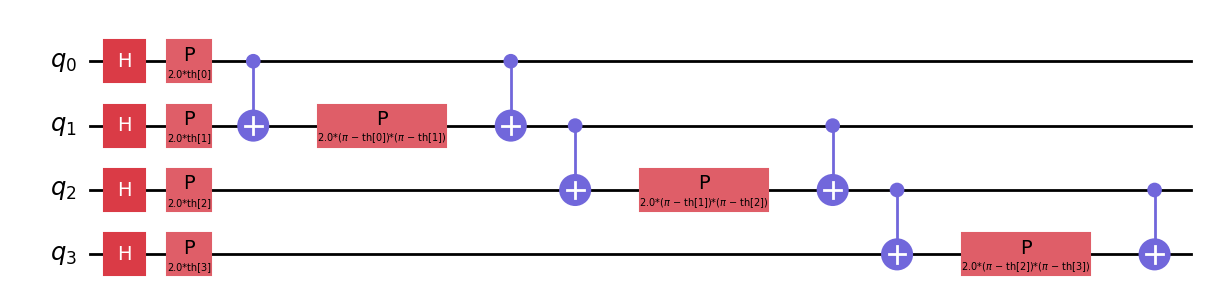}
    \caption{Complete ZZ Feature map circuit with linear entanglement circuit for 3 qubits.}
    \label{fig:circ-ZZ-linear}
\end{figure}

An example of a complete feature map circuit, using linear entanglement, is found in Fig. \ref{fig:circ-ZZ-linear}, which combines the overall structure explained in section \ref{sec:featuremap} and the linear entanglement routine. This circuit is generated directly from the Qiskit \texttt{QuantumCircuit}.

\bibliography{main.bib}

\end{document}